\documentclass{article}

\usepackage{graphicx}
\usepackage{rotating}
\usepackage[numbers]{natbib}
\usepackage{epstopdf}
\def \CeiO        {\hbox{C$^{18}$O}}                    
\def \meth        {\hbox{CH$_3$OH}}                     
\def \hctn        {\hbox{HC$_3$N}}                      
\def \nthp       {\hbox{N$_2$H$^+$}}                    
\def \arcsec      {\hbox{$^{\prime\prime}$}}            
\def \kms         {km s$^{-1}$}                         
\def \Msol  {\hbox{M$_{\odot}$}}                         
\def\ltsim{\lower 0.7ex\hbox{$\buildrel < \over \sim\ $}}
\def\gtsim{\lower 0.7ex\hbox{$\buildrel > \over \sim\ $}} 

\begin{document}
  
%

\begin{center}
{\Large\bf   Observations of chemical differentiation \\[1mm] in clumpy molecular clouds} \\[1cm]
{\large Jane V. Buckle$^1$, Steven D. Rodgers$^2$, Eva S. Wirstr\"om$^3$, Steven B. Charnley$^2$,\\ 
Andrew J. Markwick-Kemper$^{2,4}$, Harold M. Butner$^5$ and Shigehisa Takakuwa$^{6,7}$ \\ [3mm]}
{
$^1$ Astrophysics Group, Cavendish Laboratory, J J Thomson Avenue, Cambridge, CB3 0HE, UK.\\ 
email: j.buckle@mrao.cam.ac.uk\\ 
$^2$ {Space  Science \& Astrobiology Division, NASA Ames Research Center, 
Moffett Field, CA 94035, USA. }\\ 
$^3$ Onsala Space Observatory, Chalmers University of Technology, SE - 43992
Onsala, Sweden \\ 
$^4$ Astronomy Department, University of Virginia,  USA \\ 
$^5$ Joint Astronomy Centre, 660 North A'ohoku Place, Hilo, HI 96720, USA \\ 
$^6$Harvard-Smithsonian Center for Astrophysics,  Submillimeter
Array Project,\\ 
645 North A'ohoku, Hilo, HI 96720 USA. \\
$^7$ ALMA Office, National Astronomical Observatory of Japan, Tokyo, 181-8588, Japan
}

\end{center}
\noindent  
\hrule
 \vspace{0.5cm}
\noindent We have extensively mapped a sample of dense molecular clouds
(L1512, TMC-1C, L1262, Per 7, L1389, L1251E) in lines of 
 \hctn, \meth, SO and \CeiO. We demonstrate that a high degree of
 chemical differentiation is present in all of the observed clouds. We
 analyse the molecular maps for each cloud, demonstrating a systematic
 chemical differentiation across the sample, which we relate to the
 evolutionary state of the cloud.  We relate our observations to the
 cloud physical, kinematical and evolutionary properties, and also
 compare them to the predictions of simple chemical
 models. The implications of this work for understanding the origin of
 the clumpy structures and chemical differentiation observed in dense
 clouds are discussed.
\vspace{0.5cm}
\hrule
\section{Introduction}
Low-mass stars form in dense cloud cores. Understanding how stars form
  thus requires detailed knowledge of the physical and chemical
  evolution of molecular clouds.  The classic picture, in which star
  formation is magnetically-dominated and mediated by ambipolar
  diffusion \cite[e.g.][]{shu87}, posits long molecular cloud lifetimes
  ($\sim 10^7$ years).  However, there is an increasing realisation
  that dissipation of turbulence, driven externally at large scales,
  plays a vital role \cite[e.g.][]{maclow04}. In this case, star
  formation is dynamic and rapid, and molecular cloud lifetimes are
  short ($\sim 10^5-10^6$ years).  Dense cloud chemistry will follow
  different evolutionary paths depending upon the lifetime of
  molecular clouds, and hence on the time-scale of low-mass star
  formation \cite[e.g.][]{hartmann01}.  One may also expect that the
  spatial distribution of molecules should also reflect the underlying
  cloud dynamics.  Observations of dense interstellar clouds in a
  variety of molecular tracers show that they contain a distribution
  of dense gaseous structures \cite{evans99}.  These structures
  exhibit a size distribution ranging from that of dense cores
  \cite[$\sim 0.05-0.1$ pc,][]{benson89}, to clumps \cite[$\sim
  0.01-0.05$ pc,][]{peng98,takakuwa00} down to so-called small {\it
  clumpinos} \cite[$\sim 0.005$ pc,][]{takakuwa03}. The spatial
  distributions of molecules in cold, apparently quiescent, clouds do
  appear to show chemical abundance gradients, i.e. chemical
  differentiation.  Previously, the best studied sources have been the
  clouds TMC-1 and L134N \cite{swade89,pratap97,dickens00}. These
  clouds show striking differences in the location of the emission
  peaks in several molecules.  In TMC-1, for example, emission from
  the carbon chain molecules (e.g. the cyanopolyynes) is observed to
  be anticorrelated with that of ammonia and $\rm N_2H^+$ and several
  deuterated species \cite[e.g.][]{olano88,turner01,hirota01}. These
  clouds also show distinct emission peaks from other specific
  molecules, such as methanol and SO. High resolution observations
  show that chemical differentiation is present at all spatial scales
  \cite{takakuwa03,morata03,tafalla04}.  Although, in dense cores,
  depletion of CO, CS and other heavy molecules on to dust grains is
  undoubtedly important \cite{bergin02,caselli03}, the
  reason for the spatial differentiation seen amongst other molecules
  is unclear, as is its prevalence in other molecular clouds.
        
   It is possible that there is a link between the observed chemical
differentiation and the clumpy structure, and hence to the physical
origin of clumps through the cloud turbulence. There are currently two
possibilities \cite[see][]{elmegreen04a,elmegreen04b}.  First,
numerical simulations show that dissipation of externally driven
turbulence can produce structures, similar to the clumps and cores
that are observed, through the process of `turbulent fragmentation'
\cite{maclow04}.  Second, winds and outflows have a significant effect
on the structure and evolution of the surrounding protostellar
envelope, containing energies sufficient to physically disrupt the
envelope material and to alter the chemistry.  These outflows begin
very early in the star formation process, and cloud turbulence could
be excited by them \cite{reipurth01}.  In this case, the cloud
structure is regulated by the action of stars \cite[e.g.][]{norman80}.
The presence of parsec-scale flows \cite[e.g.][]{walawender05} and
wind blown bubbles \cite{quillen05} are clear observational evidence
that protostars affect their environment in this way, and that the
influence of existing star formation can have a very long reach. In
this picture, the turbulence should be a function of the local
star-formation activity, however it is unclear if enough kinetic
energy can be injected in all clouds to sustain the observed level of
turbulence.  As the nature of the turbulence should affect the
formation and evolution of clumpy structure, we might therefore expect
that there should be chemical differences between starless cores and
those containing protostars at the Class 0 and/or Class I stages
\cite{andre}, with the former being the product solely of
turbulent fragmentation.  Models have previously been developed of the
spatial and temporal chemical evolution in clumpy molecular clouds
\cite[e.g.][]{bergin02,charnley88,suzuki92,aikawa05,markwick00,charnley01,garrod05},
but with limited success.
 
To quantify the chemical effects that embedded protostars may have on
cloud structure and chemistry, we have previously mapped the Barnard 5
cloud (B5) in many molecular lines. B5 is an active region of star
formation in Perseus and contains molecular outflows, wind-blown
bubbles, and regions where the outflows appear to be interacting with
dense clumps \cite{goldsmith86,fuller91,kelly96}. Most previous
observations have focused on the region containing the Class I source,
B5 IRS1, and the molecular distributions over the rest of the cloud
were previously poorly known.  These observations revealed a high
degree of spatial differentiation, with similarities to the molecular
distributions in TMC-1 \cite{charnley05}, but also to some other
starless cores in Taurus \cite{butner05}.  Comparison with
dynamical-chemical models of B5 \cite{charnley88}, based on the Norman
\& Silk \cite{norman80} picture, appear to indicate that molecular
ices are continually being formed and destroyed
\cite{takakuwa00,butner05}, but that processes not included in these
models must be responsible.  However, B5 is only one such source; to
constrain any chemical models multi-molecule maps of more sources are
needed, for example, to see if similar molecular differentiation
occurs in dense clouds with varying degrees of star formation activity
and protostellar evolution.  We have therefore increased our sample to
observe dense core chemistry in different natal environments: starless
cores, and those containing young stellar objects (YSOs) at the Class
0 and Class I phases.  The sources were selected from the sample
mapped by Caselli et al. \cite{caselli02} in N$_2$H$^+$ at 0.063 \kms,
54\arcsec\ resolution; lower resolution ammonia maps also exist
\cite{benson89}.  Caselli et al. \cite{caselli02} found that the
position of the N$_2$H$^+$ peak emission tended to be close to the
position of peak $\rm NH_3$ emission but generally not coincident with
that of the star.  From our B5 observations, we found that several
molecules are strongly anticorrelated with $\rm N_2H^+$ and $\rm NH_3$
emission at small scales, as well as with each other (e.g. methanol
and the carbon chains).  Hence, the N$_2$H$^+$ maps of Caselli et
al. \cite{caselli02} can be useful to predict where other molecules
could be expected to peak in a cloud, assuming the chemical
differentiation has the same underlying cause.

The six cores we observed have a range of star-forming activity (see
Table \ref{table-cores}). TMC-1C is a starless core, with a region of
diffuse sub-mm emission to the south, which extends in a ridge beyond
our map to the north \cite{schnee05}. This cloud has previously been
mapped in carbon chain molecules \cite{cernicharo84} and in methanol
and HC$^{13}$O$^+$ \cite{takakuwa03}.  L1389 is associated with an
IRAS source, IRAS 04005+5647, although it has been classified as
pre-stellar, or an extremely young protostellar source
\cite{laundhardt97}. L1262 contains both a stellar core, IRAS
23238+7401, which drives a protostellar outflow, and a starless core
\cite{shirley00}. L1251E contains a YSO, IRAS 22385+7457, which drives
an outflow, as well as two infrared sources likely to be T Tauri stars
\cite{kun93}. There are also several infrared sources in the immediate
vicinity of our map. L1512 is a starless core, containing a
north-south ridge of diffuse sub-mm emission \cite{shirley00}. Per7
contains a sub-mm peak in addition to the Class I source, IRAS
03295+3050 \cite{walawender05}.

\begin{table}[t]
\caption{\label{table-cores} Core Properties}
\centering
\begin{tabular}{llllrr}
\hline
Core & RA(1950) & DEC(1950) & Class & $V_{\rm lsr}$ & D~~ \\
&&&& (km s$^{-1}$) & (pc) \\
\hline
Per7&03:29:39.5&+30:49:50&I & 6.81 & 350 \\
L1389&04:00:38.0&+56:47:59&prestellar/0 & -4.65 & 600 \\
TMC-1C&04:38:34.5&+25:55:00&prestellar & 5.27 & 140 \\
L1512&05:00:54.4&+32:39:00&prestellar & 7.11 & 140 \\
L1251E&22:38:36.4&+74:55:50&mixed & -3.93 & 350 \\
L1262&23:23:32.2&+74:01:45&0 & 4.11 & 200 \\
\hline
\end{tabular}\\
\end{table}

We chose to observe four molecular species (see Table 2). The isotope
$\rm C^{18}O$ serves to trace much of the low density ($\sim 10^3~\rm
cm^{-3}$) material in the cloud.  Methanol is formed solely on grains
in dense clouds \cite[e.g.][]{ehrenfreund00}.  Once desorbed into the gas, methanol molecules only have a finite lifetime until they are destroyed.  Hence, the  CH$_3$OH distribution provides an excellent observational measure of the degree of gas-grain cycling in clumpy clouds.
 Cores showing weak methanol emission will tend to be older since any
CH$_3$OH liberated previously will gradually re-accrete back onto the
dust.  As we discovered a new 'cyanopolyyne-peak' position in B5,
similar to that in TMC-1 \cite{charnley05}, we also mapped the
cyanoacetylene emission in these clouds.  The formation and
destruction of cyanopolyynes can provide information on those
molecules returned to the gas, and also define chemical time-scales in
dynamically-evolving clouds \cite{markwick00}.  Markwick et
al. \cite{markwick00} concluded that such chemically interesting
positions should appear to be offset from the position of any YSO
present; as well as from the peak  ammonia emission.
The detection of these organic molecule `hotspots' is thus only
possible by making large-scale $ {\rm HC_3N}$ maps, to include
positions distinct from the nominal (0,0) positions of previous
searches \cite[cf.][]{suzuki92}. The \hctn\ transition chosen traces
gas at densities above $\sim 10^5~\rm cm^{-3}$. Finally, to explain
the large number of S-bearing molecules detected in dense clouds,
chemical models infer that elemental sulphur is only modestly depleted
in dense clouds, relative to more refractory elements
\cite{ruffle99}. Interstellar sulphur chemistry involves primarily
neutral reactions, and so it defines yet another, albeit slow,
chemical time-scale. Shock chemistry or grain-surface catalysis have
been proposed to be important for S chemistry
\cite[]{charnley01,millar90}; the latter mechanism is supported by the
detection of OCS in interstellar ices \cite{palumbo97}. We therefore
mapped the $J_{\rm K}=3_2-2_1$  line of SO.

\begin{table}[t]
\caption{\label{table-mols} Molecular Line Properties}
\centering
\begin{tabular}{llrrrr}
\hline
Molecule & Transition & Frequency & E$_{\rm upper}$ & Einstein A & n$_{\rm crit}$ \\
&& (GHz) &(K) & (s$^{-1}$)& (cm$^{-3}$)\\
\hline
\CeiO&1$\rightarrow$0&109.782&5.27&	    6.30 $\times10^{-08}$& $2\times10^3$  \\     
SO  &2$_3\rightarrow$1$_2$&99.300&9.22&	    1.10 $\times10^{-05}$& $3\times10^5$ \\     
\hctn&10$\rightarrow$9&90.979&24.01&	    5.81 $\times10^{-05}$& $7\times10^5$  \\     
\meth&2$_1\rightarrow$1$_1$(E)&96.739&12.55&2.56 $\times10^{-06}$& $1\times10^5$  \\     	
&2$_0\rightarrow$1$_0$(A$^+$)&96.741&6.97&  3.38 $\times10^{-06}$& $5\times10^4$ \\     	
&2$_0\rightarrow$1$_0$(E)&96.744&20.11&	    3.41 $\times10^{-06}$& $4\times10^4$ \\      	
&2$_1\rightarrow$1$_1$(E)&96.755&28.04&	    2.62 $\times10^{-06}$& $5\times10^4$ \\      	
\hline
\end{tabular}\\
\end{table}

The plan of the paper is as follows. In Section 2 we describe the
observations and how the data analysis was performed.  In Section 3,
we detail the chemical morphology of each source and note the general
trends we see in chemical differentiation. Simple chemical models of
the chemical evolution in clumpy media are presented and compared to
the observational data in Section 4. Conclusions from this study are
given in Section 5.

\section{Observations, Data Reduction and Analysis}

Single dish maps of \hctn, SO, \meth\ and \CeiO\ transitions at 3mm
were made using the Onsala 20m telescope, Sweden, in April 2005. With
the exception of \meth, the spectra were observed in
frequency-switched mode, using 12.5 kHz channels to provide a velocity
resolution of $\sim$0.04 \kms.  \meth\ was observed with the same
resolution, but in position-switched mode due to the number of
transitions in the band. The 20m beam size at these frequencies is
approximately 40$\arcsec$, and the main beam efficiency, $\eta_{mb}$,
is 0.46.  Table \ref{table-mols} lists molecular line details for the
transitions that we observed.

The cores observed were taken from Caselli et al. \cite{caselli02},
and include cores at a range of evolutionary stages, from prestellar
to Class I (Table \ref{table-cores}). The areas mapped cover the
N$_2$H$^+$ emission regions within each core \cite{caselli02},
ranging in size from 3--20 sq. arcmin. The typical RMS noise level in
the maps is 0.1--0.3~K. Contour plots and channel maps are shown in
Figs \ref{fig-contour-1}--\ref{fig-contour-6}.

Data reduction was carried out using the radio data reduction packages
XS (bespoke software for the Onsala 20m telescope), CLASS and SPECX.
Single Gaussian fits provide a good fit to the line profiles. For
\meth, the transitions detected in each spectra were fit with
dependent Gaussians. The velocity and linewidth of each transition was
fixed with respect to the 2$_0 \rightarrow$1$_0$($A^+$) transition at
96.741~GHz, since the separation of the K transitions is fixed by the
structure of the molecule. The results of the fitting procedures are
given in Table \ref{t_fits}, and for \meth, values are given for the
2$_0 \rightarrow$1$_0$(A$^+$) transition. For lines that were not
detected, the 3$\sigma$ error estimate is given as the upper limit to
the integrated intensity.

Since we have observed single transitions in \CeiO, SO and \hctn, and
detect only one or two \meth\ transitions in a large fraction of the
spectra, we determine a lower limit to the column density
\cite{thompson99}. In this procedure, the rotation temperature can be
approximated as $\rm T_{rot}=\frac{2E_u}{3k}$ for non-linear molecules
(\meth), and as $\rm T_{rot}=\frac{E_u}{k}$ for linear molecules
(\CeiO, SO and \hctn). A lower limit to the column density can then be
calculated as:
 \begin{eqnarray}
 \rm N_{min}&= \rm\frac{ 8 \pi k \nu^2 }{h c^3 A_{ul} g_u}~\int \rm T_{mb}dv~Q(\frac{2E_u}{3k})e^{3/2} & {\rm non-linear}\\
 \rm N_{min}&= \rm\frac{ 8 \pi k \nu^2 }{h c^3 A_{ul} g_u}~\int \rm T_{mb}dv~Q(\frac{E_u}{k})e & {\rm linear}
 \end{eqnarray}
where $\rm T_{mb}$ has been calculated from $\rm T_{A}^*$ using
$\eta_{mb}$, partition functions (Q[T]) and statistical weights
(g$_u$) have been taken from the JPL molecular line database
\cite{pickett98}, energy levels and Einstein A values (A$_{ul}$) have been
taken from the Leiden Atomic and Molecular Database \cite{schoier05}
and Cragg et al. \cite{cragg93}, listed in Table~\ref{table-mols}. For
\meth, the partition function Q(T$_{\rm rot}$)=1.28T$_{\rm rot}^{1.5}$
was adopted from Menten et al. \cite{menten86}, and takes into account
equal populations of A and E species \cite{kalenskii97}.

For sources where we have detected 3 or 4 \meth\ transitions, we have
also carried out a rotation diagram analysis. The resulting column
densities generally agree to within a factor of a few. However, as we
 have detected only one strong transitions in each A and E
species towards most positions, we have based our \meth\ analysis in
this paper upon the 2$_0 \rightarrow$~1$_0$(A$^+$) transition, in the
same manner as for the other molecules. This provides a coherent
analysis of this data that we can compare with simple chemical models.

From our \CeiO\ observations, we can then derive abundances relative
to H$_2$ following Frerking et al. \cite{frerking82} for N(\CeiO)~$\geq$~3
$\times 10^{14}$ ~cm$^{-2}$:
 \begin{equation}
 N(H_2) = \left[\frac{N(\CeiO)}{1.7\times10^{14}}+1.3\right]\times 10^{21}~\rm cm^{-2}
 \end{equation}
We derive H$_2$ column densities $\sim 2\times 10^{22}$~cm$^{-2}$
within the clumpy regions of our data, leading to abundances on the
order of $x$(\meth)~$\sim$~2~$\times$~10$^{-8}$,
$x$(\hctn)~$\sim$~8~$\times$~10$^{-11}$ and
$x$(SO)~$\sim$~4~$\times$~10$^{-10}$ in the molecular clumps, or `hotspots'.

\section{Molecular Morphology }

\begin{figure*}[t]
\includegraphics[width=15cm]{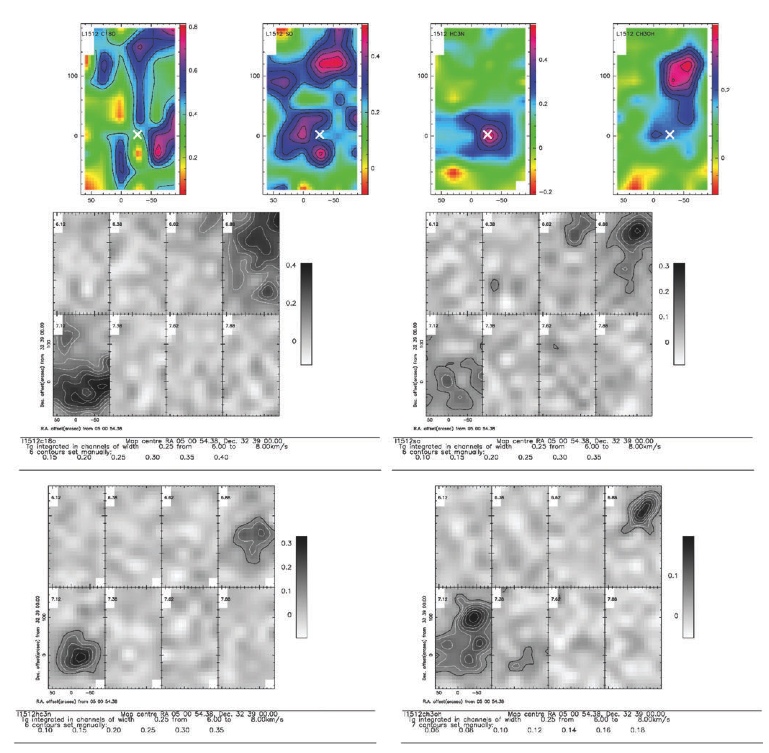}
\label{fig-contour-1} 
\caption{L1512. Top: Contour maps of
integrated intensity, from left: C$^{18}$O, SO, HC$_3$N and CH$_3$OH.
Contours are 50\%--100\% in 10\% intervals of the peak integrated
intensity in K \kms, and 1$\sigma$ rms is $\sim$ 0.04 K \kms. A cross
marks the N$_2$H$^+$ peak \cite{caselli02}. Bottom: Channel maps of
integrated intensity. Contours levels are marked on each
plot. Clockwise from top left: C$^{18}$O, SO, CH$_3$OH and HC$_3$N.}
\end{figure*}

\begin{figure*}[t]
\includegraphics[width=15cm]{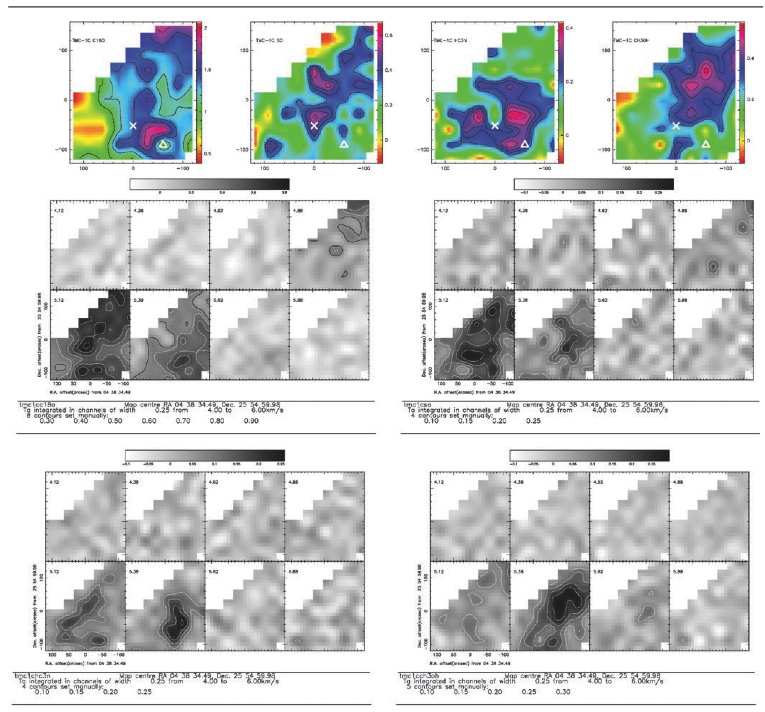}
\label{fig-contour-2}
\caption{TMC-1C.  As Fig. \ref{fig-contour-1}. A triangle marks
the approximate centre of the diffuse sub-mm emission to the south,
which extends slightly to the east, and to the north beyond our map
\cite{schnee05}. A cross marks the N$_2$H$^+$ peak
\cite{caselli02}.}
\end{figure*}

\begin{figure*}
\includegraphics[width=15cm]{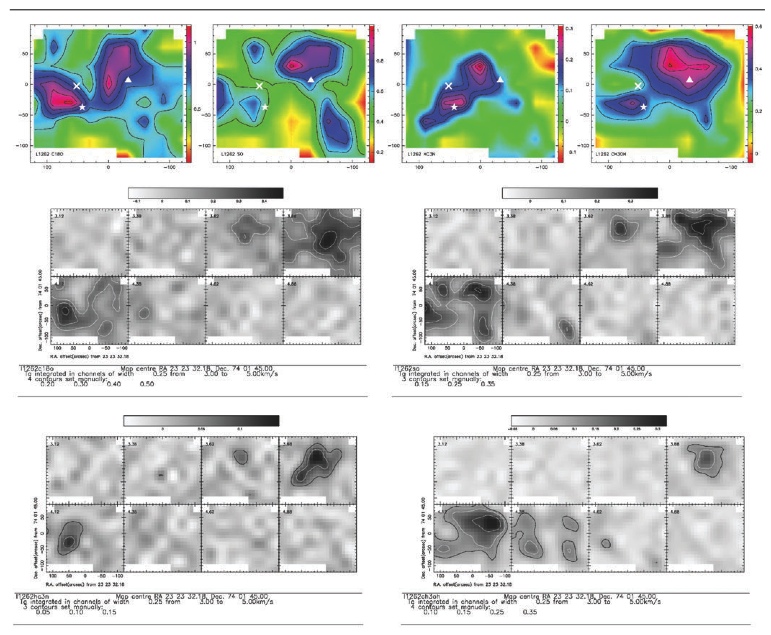}
\label{fig-contour-3}
\caption{L1262.  As
Fig. \ref{fig-contour-1}. The IRAS source is marked with a star, and
 sub-mm peaks with triangles \cite{shirley00}. A cross marks the
N$_2$H$^+$ peak \cite{caselli02}. }
\end{figure*}

\begin{figure*}[t]
\includegraphics[width=15cm]{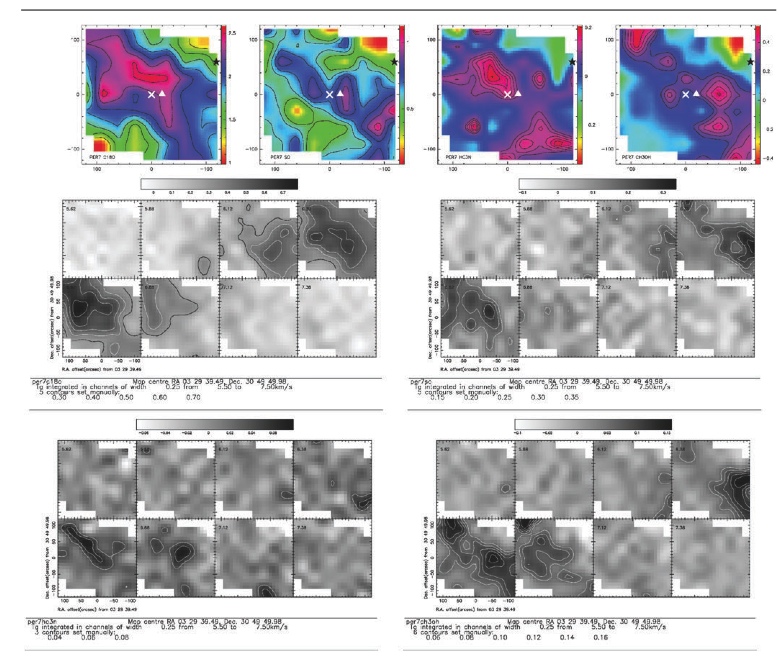}
\label{fig-contour-4}
\caption{Per7. As Fig. \ref{fig-contour-1}.  A
  star marks the IRAS source, a triangle marks the sub-mm peak
  \cite{walawender05}, and a cross marks the N$_2$H$^+$ peak
  \cite{caselli02}.}
\end{figure*}

\begin{figure*}[t]
\includegraphics[width=15cm]{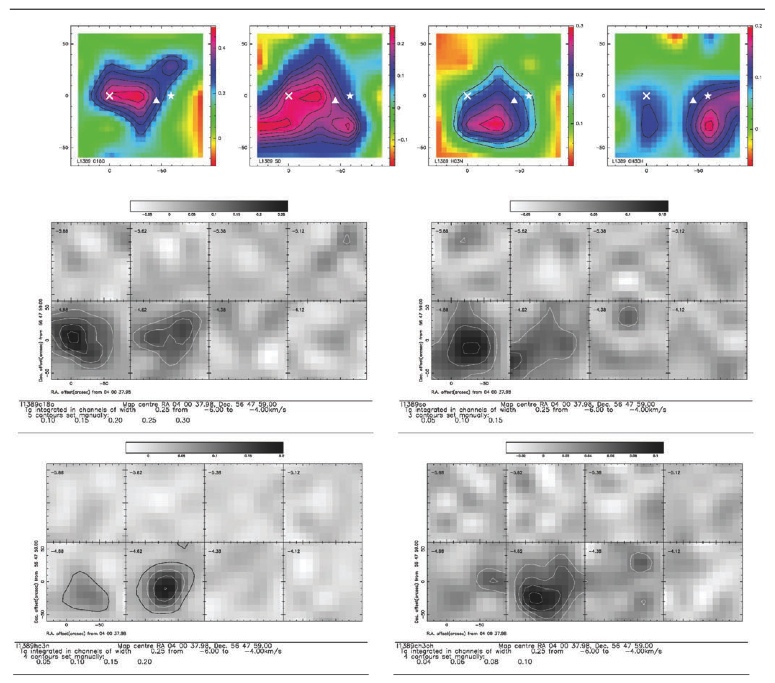}
\label{fig-contour-5}
\caption{L1389. As Fig. \ref{fig-contour-1}. A star
marks the IRAS source, a triangle marks the sub-mm
peak \cite{laundhardt97}, and a cross marks the N$_2$H$^+$ peak
\cite{caselli02}.}
\end{figure*}

\begin{figure*}[t]
\includegraphics[width=15cm]{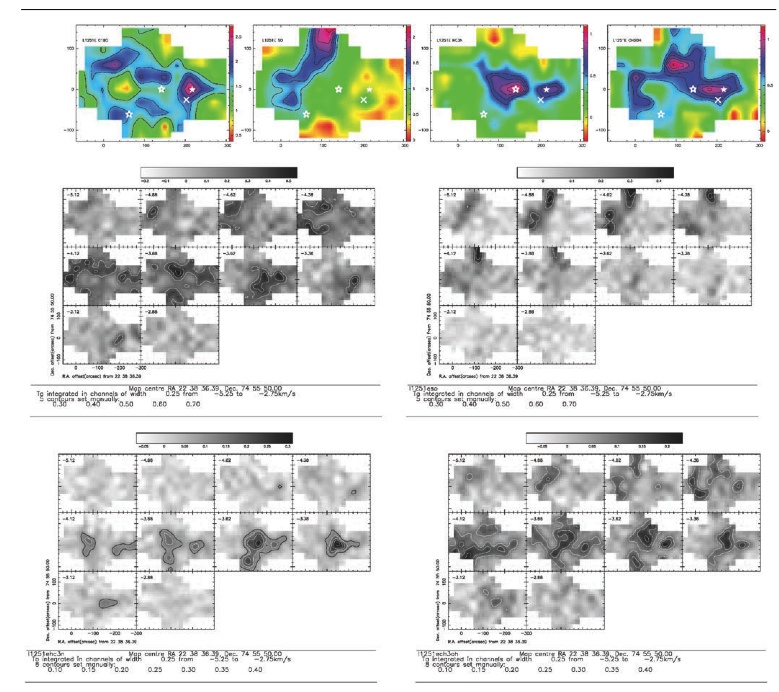}
\label{fig-contour-6} 
\caption{L1251E.  As
 Fig. \ref{fig-contour-1}. The IRAS source is marked with a filled star, and
 unidentified IR sources with empty stars \cite{kun93}. A cross marks the
N$_2$H$^+$ peak \cite{caselli02}.}
\end{figure*}

 We now describe the observed chemical morphology. We first discuss
 the trends between the four molecules in each individual source,
 focusing on the emission present in the position-velocity channel
 maps (Fig.~\ref{fig-contour-1}--\ref{fig-contour-6}).  We then
 summarise the apparent general trends.
 
\subsection{Individual Sources}

\subsubsection*{L1512}

L1512 shows several extended clumps and ridges of emission in
\CeiO. The southern part of the map shows two clumps
((-60\arcsec,-30\arcsec) and (0\arcsec,-30\arcsec)) which surround the
compact \hctn\ clump (-30\arcsec,-30\arcsec). Emission from SO is also
seen in extended ridges and clumps. The strongest clump is associated
with the single \hctn\ clump (-30\arcsec,-30\arcsec), while the second
strong clump is associated  with the \meth\ peak to the north
(-60\arcsec,120\arcsec).

We see SO emission at 6.62 \kms\ with none apparent in either \meth\
or \hctn.  The strongest SO emission clump appear in gas in which
\meth\ is present, at 6.88 \kms, the SO and \meth\ emission peaks are
offset by approximately 30\arcsec.  The methanol peak is at 7.12 \kms,
where most of the weaker clumpy emission is found; at this velocity,
there is no corresponding spatial SO emission in this region.  There
are a couple of weak \meth\ emission clumps present to the south, and
these are embedded in the same gas as two SO clumps.  An \hctn\ clump
shows up at -6.88 \kms\ that is spatially distinct from the SO and
\meth\ clumps but partially overlaps the weaker extended SO
emission. The main \hctn\ clump is at 7.12 \kms, overlaps the extended
SO and \meth\ there, but is distinct from the weak SO and \meth\
clumps, and is located about 100\arcsec\ from the major SO-\meth\
clumps.  In fact, we find that both these \hctn\ clumps coincide with
minima in the \CeiO\ distributions, as well as with the $\rm N_2H^+$
peak ($V_{\rm LSR}$ = 5.27 \kms), perhaps indicating that molecular
depletion is important at these positions. However, the fact that a
methanol clump is present at 7.12 \kms\ in a region with little/no
\CeiO\ emission somewhat confuses this interpretation. \CeiO\ emission
peaks are anti-correlated with emission peaks in all other molecules.


In summary, we see gas with only SO, gas with strong SO and \meth\
emission intermingled but with distinct emission peaks (i.e. clumps),
as well as gas with strong \meth\ emission and no/weak SO.  The \hctn\
clumps are markedly anticorrelated with SO and \meth\ clumps in
position and are correlated with \CeiO\ troughs. These could be
cyanopolyyne late-time depletion peaks \cite{ruffle97}.

\subsubsection*{TMC-1C}

TMC-1C shows several clumps of emission in \CeiO\ extending in a
north-south ridge. The peak of emission is near the centre of the
diffuse sub-mm emission (-30\arcsec,-60\arcsec), with several smaller
clumps leading north to a more extended emission region at the edge of
our map (-90\arcsec,120\arcsec). \hctn\ emission is confined to the
southern half of our map, where it appears in a ring, part of which
overlaps the \CeiO\ clump. SO emission appears in a relatively narrow
ridge of emission extending to the north-west. Emission from
\meth\ has a similar morphology to emission from SO, but is more
extended, and peaks further to the north.

In this source SO displays very clumpy emission at 5.12 and 5.38 \kms;
there is only weak/no \meth\ emission at 5.12 \kms, where most of the
SO molecular clumps are.  At 5.38 \kms, \meth\ has 3 clumps and there
is evidence for coincident weak SO emission.  There is an SO clump
coincident with a minimum in \CeiO\ emission; this is approximately
the location of $\rm N_2H^+$ emission maximum \cite{caselli02}, and
suggests a depletion zone here.  This is curious: in L1512 we have
\hctn\ correlated with a possible depletion zone, whereas here it is
SO and there is only weak/no \hctn\ present - or at least tending to
avoid it.  The $\rm N_2H^+$ peak is at $V_{\rm LSR}$ of 5.27 \kms, at
a similar velocity to emission from \CeiO, \meth\ and \hctn. The
velocity of SO emission at this position is blue-shifted by 0.1 \kms,
which presumably means that the depleted material does not contain the
SO peak.  The structure of the \CeiO, \hctn\ and SO emission could
be interpreted as that of a clumpy, fragmented torus in 3D. In this
scenario, the SO emission is actually coming from a chemically
differentiated fragment of the torus/ring.  The \hctn\ peaks are
spatially distinct from the \meth\ clumps at 5.38 \kms\ but there is
weak SO emission in them; an exception is the SO clump near
(0,+60\arcsec) which appears to have no \hctn. Again, \CeiO\ emission
peaks are anti-correlated with emission peaks in all other molecules.


Thus, in the SO clumps, methanol is either deficient or absent. To
some degree, SO and \hctn\ appear to be intermingled, although the
clumps/peaks are physically distinct, and one SO clump has no \hctn.
The \meth\ clumps probably have some SO in them but are spatially
distinct from the \hctn\ distribution.  \CeiO\ emission minima
correspond to SO maximum (0,-60\arcsec) and to \hctn\ maximum
(-60,-90).

\subsubsection*{L1262}

Emission from \CeiO\ towards L1262 is seen in two strong clumps, one
just to the east of of the IRAS source (0\arcsec,0\arcsec), and one
just to the east of the sub-mm source (60\arcsec,-30\arcsec). Emission
from \hctn\ is seen in a clumped ridge, with one clump, like that of
\CeiO, to the east of the sub-mm peak, and one associated with the
IRAS source, west of the \CeiO\ clump (30\arcsec,-30\arcsec). SO
emission peaks in a clump to the north of the sub-mm peak
(0\arcsec,30\arcsec), and extends in a clumpy ridge to the north and
south. A further clump is seen to the west, at the edge of our map,
where the \CeiO\ emission also extends. Emission from \meth\ is seen in
an extended clump associated with the sub-mm source
(0\arcsec,30\arcsec), and also in a compact clump to the east of the
IRAS source (60\arcsec,-30\arcsec). The main SO and \meth\ clumps are
co-incident with the \hctn\ clump that is near the sub-mm
peak.

There is an SO clump at 3.62 \kms\ that appears to grow and merge into
 the major SO clump at 3.88 \kms, where \meth\ emission becomes more
 evident.  At 4.12 \kms, the SO emission clearly shows up as $\sim$ 4
 clumps.  There are 2 \meth\ clumps and these are evident at 4.12 and
 4.38 \kms. In both cases, the \meth\ appears to be mixed with the
 enhanced SO emission, although at 4.38 \kms\ the SO emission is
 absent.  The strongest SO and \meth\ emission are to the north - the
 clumps/peaks are spatially distinct and should have different
 SO/\meth\ ratios. The methanol clump at 4.38 \kms\ is almost `pure'
 \meth\ - emission from other molecules is weak/absent. There are 2
 \hctn\ clumps with the strongest to the north at 3.88 \kms. This
 clump is associated with the strongest SO peak. Only the weaker
 \hctn\ clump to the south is also evident at 4.12 \kms\ which is
 located across a region of weak/absent SO and \meth\ emission.  There
 are two \CeiO\ clumps, which are distinct both spatially and
 kinematically. Through both of these, we see a kinematic sequence of
 \CeiO\ and  SO, then \CeiO\ and SO plus \meth, then just \meth\ as we move
 from the blue to the red in velocity channels, although the spatial
 peak in emission from these molecules are offset from each other.


In L1262, we detected 4 SO clumps. The strongest one  also
contains quite strong \hctn\ and weak \meth\ emission at 3.88 \kms;
however, at 4.12 \kms\ the \meth\ clump grows stronger and more
distinct while the \hctn\ emission disappears.  This is similar to
that seen previously, where the strongest SO and \meth\ clumps tend to
be in gas where the molecules coexist. We note that the methanol
clumps are spatially anticorrelated with the positions of the \hctn\
ones, and that the weaker \meth\ and \hctn\ clumps have little SO.
However, the positional offset of the emission peaks of each molecule
is only marginal ($\sim$0.5 beamwidth).

 The SO, \meth\ and \hctn\ peaks are all anticorrelated with \CeiO\ peaks. 
 These molecules also tend to avoid the region
around the $\rm N_2H^+$ emission peak (at $V_{\rm LSR}$ = 4.11 \kms), suggesting that molecular depletion on grains could be influencing the
observed morphology. 
There may be a kinematical sequence through the clumpy structures
around (-60, 0). 

\subsubsection*{Per 7}

Per7 shows several regions of clumpy, asymmetric emission from all
four molecules, most of it contained in ridge extending to the
north-east and south-west from the sub-mm peak. Emission from
\CeiO\ is relatively bright towards this source, while emission from
\hctn, \meth\ and SO is fairly weak. \CeiO\ peaks in an extended
region to the NE of the sub-mm peak (around
(30\arcsec,30\arcsec)). \hctn\ peaks in the same place, but emission
is much more compact, with a second emission peak to the SW
(-90\arcsec,-60\arcsec). SO appears in a series of clumps along the
NE-SW ridge, with the brighter clumps those to the SW
(-90\arcsec,-60\arcsec). \meth\ also appears in a series of clumps
along the NE-SW ridge, but the brightest clump is to the far NE
(90\arcsec,90\arcsec), where no SO emission is seen. Another peak
occurs in between the two strongest SO emission clumps
(-60\arcsec,0\arcsec).

Very clumpy structure is evident in emission from SO.  At 6.38
\kms\ there are 2 SO clumps to the north with no/little \meth\ or
\hctn\ emission. However, the peak emission clumps of SO and
\meth\ are in the same gas, but physically distinct (there is also a
possible/weak \hctn\ clump at this velocity).  At 6.62 \kms, SO and
\meth\ show several clumps in gas where they are comingled but with
the peaks positionally offset from each other. There are 3 major
\meth\ clumps and it seems that the northern one contains little or no
SO and definitely no \hctn. At 6.62 \kms, there is one clear
\hctn\ clump that actually sits in the \meth\ minimum. In fact, once
again, the \hctn\ and \meth\ emission is generally anticorrelated. The
\CeiO\ does not show compact peaks in the contour map, but is also
anticorrelated with \hctn\ and SO.  All the molecules avoid the
associated $\rm N_2H^+$ peak.

   
In this source we again see clumps with essentially only SO present,
     as well as that the peak emission from SO and \meth\ emanates
     from gas in which they are comingled, but where their emission
     peaks/clumps are spatially distinct.  There appear to be methanol
     clumps with no \hctn\ and little or no SO. Again, there appears
     to be a chemical gradient through the clumpy structures at around
     (+100, 0): 6.38 \kms\ : SO only - 6.62 \kms\ : strong \meth\ and
     unchanged SO : 5.88 \kms\ strong \meth\ no SO.

\subsubsection*{L1389}

L1389 shows relatively compact and weak emission from all four
molecules. \CeiO\ peaks in a clump to the
east of the IRAS source (-30\arcsec,0\arcsec). \hctn\ peaks in a clump
south of the \CeiO\ emission. Emission from SO also peaks to the east
of the IRAS source (-30\arcsec,-30\arcsec), but is more
extended. \meth\ emission appears in a strong clump to the south of
the IRAS source, extending to the west (-60\arcsec,-30\arcsec).

Chemically, this  is a morphologically simple cloud. The SO shows a single clump at -4.88
\kms\ with weak, diffuse emission at -4.62 \kms; this clump appears to
have very weak \hctn\ emission but no \meth.  The \meth\ clump is at
-4.62 \kms\ and this corresponds to a region of relatively weaker SO
emission.  The \hctn\ clump appears at -4.62 \kms\ and and is
spatially distinct from the SO peak in velocity and the \meth\ peak in
position.
 
   
 So, there is a clump in which SO is prevalent, although it does
  contain some \hctn.  A distinct \hctn-\meth\ anticorrelation is also
  evident.  All the molecular clumps (positions of peak emission) are located away
  from   $\rm N_2H^+$ peak, supporting the idea that molecular depletion is important here.

\subsubsection*{L1251E}

Towards L1251E, emission from \CeiO\ is seen in several small clumps,
the strongest of which is associated with the IRAS source
(-210\arcsec,0\arcsec). Two clumps are also seen to the north and
south of an infrared source, either side of a clump of \hctn\ emission
(-150\arcsec,0\arcsec). \hctn\ emission is seen in a second clump
associated with the IRAS source (-210\arcsec,0\arcsec). SO emission is
seen in a ridge of emission extending to the north
(-120\arcsec,120\arcsec), not clearly associated with emission from
any of the other molecules, nor any of the infra-red/IRAS sources. The
clumps near the second infra-red source are associated with the weaker
\meth\ clumps. \meth\ emission is seen strongly peaked in a clump
associated with both the IRAS source and the first infrared source
(-180\arcsec,0\arcsec), with a second peak to the north, just south of
the main SO clump (-90\arcsec,60\arcsec).

The strong SO emission clump to the north, evident from -4.88 to 4.12
  \kms, contains little or no \meth, no \hctn, and is unconnected to
  the level of \CeiO\ emission present. The \meth\ emission is
  extended and fragmented and three main clumps are evident. There are
  2 \hctn\ clumps present. The \hctn\ peak/clump at (-120, 0) appears
  quite extended, shows weak \meth\ emission but none from SO.
     

In summary, there is an SO clump that contains a small amount of
\hctn\ emission and only a small amount of \meth\ emission.  The \hctn\
- \meth\ peaks are anti-correlated to the east, where the \hctn\ peak
appears in a region with no \meth\ emission. To the west, \hctn\ and
\meth\ emission are comingled, but the peaks are offset.

\subsection{General Trends - Summary }
Several trends in the chemical differentiation are evident in our
maps. The morphology throughout our sample ranges from one or two
isolated clumps detected in each molecules, to extended emission
containing a strongly-emitting clump and a several weaker ones.

\subsubsection*{\it  SO and CH$_3$OH  Clumps}  

We find clumps of gas that contain significant emission only from SO
 (e.g. L1389, L1251E, PER 7, TMC-1C), with only one of these showing
 significant emission from \hctn\ (L1262). There is evidence that
 some SO emission maxima could correspond to depleted regions
 (TMC-1C). Strong extended SO emission is also evident where \meth\
 emission becomes stronger; in this case we find that the dominant
 \meth\ and SO clumps can exist in the same gas but that they are
 always offset from each other.  On the other hand, there are some
 methanol clumps from which SO emission appears to be absent. Based on
 comparing maps in velocity space, there is some kinematical evidence
 (e.g.  in L1262, PER 7, L1512) for well-defined emission sequence in
 which there are : regions with only SO, then a region with SO and
 \meth\ intermingled - then a region with only \meth, suggesting we could be
 looking down a cylinder/filament.

\subsubsection*{\it The HC$_3$N Clumps} 

We find that the \hctn\ clumps are {\it always} spatially
 anticorrelated with the \meth\ clumps.  Some of these clumps can
 contain some very weak \meth\ emission, and moderately strong SO
 emission (L1262), but there are others where the \hctn\ is the sole
 molecule present (L1251E).

\subsubsection*{\it Depletion: C$^{18}$O Distribution \&  N$_2$H$^{+}$ Cores}
 
$\CeiO$ emission is tracing the less dense gas towards these sources,
since the critical density of the transition we have observed is 1 or
2 orders of magnitude less than the transitions we observed in SO,
\hctn\ and \meth. The emission generally has several clumps embedded
within an extended, diffuse region, and emission from spatially
distinct regions is generally seen to be kinematically distinct as
well. For example, towards L1512, emission to the north-west is seen
at 6.88 \kms, while emission to the south-east is seen at
7.12 \kms. There is a clear anti-correlation between emission from
\CeiO\ and all other molecules towards the youngest sources, those
classified as prestellar cores. 
Towards the older sources, where protostars have formed,
emission becomes more intermingled, and \CeiO\ emission can be seen
associated with SO and \meth\ peaks, although the spatial positions are
offset (e.g. L1262).

 Another population of objects implicitly present in our maps are the
N$_2$H$^{+}$ cores mapped by Caselli et al. \cite{caselli02}. These are probably
the most stable, longest lived cores in the clouds, apart from the
ones that apparently harbour protostars, and are almost certainly
sites of extensive molecular depletion onto grains. All molecules in
our sample tend to avoid these positions.  An exception is in L1512
where the \hctn\ clump is coincident with the N$_2$H$^{+}$ peak in
both position and velocity.


\section{Chemical Differentiation in a Clumpy Medium: Theory}

We conclude that we are observing chemical morphologies which are
produced by a time-dependent gas-grain chemistry which is evolving in
a turbulent dense medium. Clumps form and dissipate, perhaps
coalescing, and long-lived stable structures (i.e.\ dense cores) are
produced in which star formation occurs. In this section we show how
simple chemical models can help explain these observations.

Clumps are formed out of the interclump medium (${ n _{\rm H} } \sim
10^3 \rm cm^{-3}$), perhaps by compression in shock waves \cite{klessen01},
 or in MHD waves \cite{garrod05}. Our observations indicate that molecular
desorption/depletion from/onto dust grains plays an important role in determining the
chemical differentiation.   For a molecule, $X$ say, of molecular
weight $\mu _{\rm X}$, the depletion (accretion) time-scale at 10K is
  \begin{equation} \label{eqn:taccrete}
 { t_{\rm X} ~=~ {   2.2 \times 10^{9 }   {  { ~~\mu _{\rm X}^{1/2} } \over { n _{\rm H} }    }  }
 \rm ~~~ yr }
 \end{equation}
where a total grain surface area per H nucleon of $3 \times 10^{-22 }
~\rm cm^{2} $ and unit sticking efficiency have been assumed
\cite[cf.][]{brown90}.  In these dense regions (${ n _{\rm H} } \sim
10^4-10^5~{\rm cm}^{-3}$) accretion and chemical reactions are driven
more rapidly and dynamical fate of the clump becomes an issue.  It is
therefore useful to know the relevant physical time-scales that are
important for clump evolution.

%
%


\subsection{Clump Physics}
 
The fate of clumps can be related simply to parameters that can be
derived from observations. From our observational data we can obtain
estimates for the clump mass, $M$, its virial mass, $M_{\rm vir}$, and
radius $R$.  With $R$ known, the virial mass for a sphere of constant
density can be estimated from \cite{maclaren88}
  \begin{equation}
  {   M_{\rm vir}   ~\sim~    210  (\Delta v )^2_{\rm av}  R ~~~   M_\odot  }
  \end{equation}
where $(\Delta v )_{\rm av} $ is the average line-width of the clump
gas (H$_2$ and He) in \kms\ and $R$ is in parsec 
For clumps in molecular clouds, there are three possible evolutionary
tracks \cite[cf.][]{peng98}.  A clump will be stable against
gravitational collapse if ($0.5M_{\rm vir}<M< M_{\rm vir} $), whereas if it is
gravitationally unstable ($M> M_{\rm vir} $) it will collapse on the
free-fall time-scale
  \begin{equation}
  t_{\rm ff}  ~=~  {{ 4.35 \times 10^{7} } \over { n _{\rm H}^{1/2} }}
  \rm ~~~ yr
\end{equation}
If ($M<0.5M_{\rm vir} $) then the clump is gravitationally unbound and
so it will dissipate on a time-scale
  \begin{equation}
  {   t_{\rm diss}   ~\sim~   {  2R   \over   C_{\rm eff}     }    }
  \end{equation}
where $C_{\rm eff}  $ is the effective sound speed in the gas.
The fate of some clumps can also depend upon the local
environment. A cloud volume containing a number of clumps of  
filling factor $\gamma$, and line-of-sight
relative velocity $ v _{\rm rel}$ can collide and coalesce
(i.e. merge) with a time-scale
  \begin{equation}
  {   t_{\rm merge}   ~\sim~   {  2R   \over    v _{\rm rel}  }     
  \left ( {1 \over  \gamma ^{\rm 1/3}}  - 1 \right)        }
  \end{equation}
Clumps with any dynamical fate can grow when the merger time-scale is
sufficiently short; unbound clumps can do so when this is comparable
or less than the dissipation time-scale, i.e. when $ t_{\rm merge}
\ltsim t_{\rm diss}$ \cite{peng98,takakuwa03,morata03}.  If we take
the derived clump properties of the Taurus clouds (TMC-1C and L1512)
as an example, we then, following Peng et al. \cite{peng98}, we
estimate that typically $R \sim 0.02-0.03$ pc and the clump masses are
in the range $M \sim0.3-1 ~\Msol$. A detailed analysis of clump
properties is beyond the scope of this paper, however, we note that
throughout our sample we can identify clumps that are unbound, as well
as some that are gravitationally stable and others that are unstable.
Some clumps, such as \meth\ and SO clumps in TMC-1C and L1512, may
actually show substructure at higher spatial resolution. In this case,
such smaller clumps would be more likely to be unbound and, given
their proximity, perhaps more prone to interaction.

 
 

\subsection{Chemical Model}

We now show how a simple static model of clump chemical evolution,
with accretion on to dust and  selective injection
of simple mantle molecules, can explain the observations.

As discussed in section 1, methanol cannot be produced efficiently in
the gas phase, and is thought to be formed on the surfaces of dust
grains. Hence, the \meth-rich clumps in our sources must necessarily
trace clumps in which grain ice mantles have recently been liberated
into the gas phase. 
The exact mechanism that causes the desorption of
the ices is unknown but we expect that it will be connected to clump evolution. 
Markwick et al. \cite{markwick00} proposed that
the desorption is driven by grain-grain collisions, induced by MHD
waves generated by clump motions. Alternatively, Dickens et al.
\cite{dickens01} speculated that grain-grain streaming resulting from
clump collisions may be responsible.
In this paper, we make no assumptions about the dynamical  mechanism, other than
that it is  related to clump formation, and so ice mantle
liberation, if it occurs, happens at the same time that the clump is
formed.  This is obviously a simplification since there must be a period of CO accretion to provide the
 precursor ices for \meth\ production.
We assume that a clump forms
instantaneously from the interclump medium,  and
simply follow the chemistry until all the molecules are condensed on
grains.    
 In reality, the chemical evolution will be sensitive to the dynamical evolution 
 of the clump (e.g. whether it is collapsing, coalescing or dissipating).   We do not consider the clump physical evolution explicitly in this static  model;  this is discussed in detail elsewhere
\cite{rodgers05}.


Based on the critical densities of the observed molecular tracers, we
adopt a value of $n _{\rm H} = 2 \times 10^5 \rm ~ cm^{-3} $ as
representative of the clump densities, which we consider to be
approximately spatially constant. This choice yields $t_{\rm ff} \sim
1 \times 10^5$ years, and $ t_{\rm diss} \sim 2 \times 10^5$ years,
and clump masses similar to those derived from the observations, viz
  \begin{equation}
  M   ~=~  0.35 M_\odot   
  \left ( {R \over 0.02 \rm ~ pc}   \right) ^3  
  \left ( {n _{\rm H}   \over   2 \times 10^5 \rm ~cm^{-3} }  \right)
  \end{equation} 
We assume a cosmic ray ionization rate of $3\times10^{-17}$~s$^{-1}$,
a temperature of 10~K, and a visual extinction of 10 magnitudes, which
is sufficiently large to render photo-processes effectively
irrelevant. Freeze-out of molecules onto grains occurs at a rate given
by eqn.~(\ref{eqn:taccrete}), except for N$_2$ which is assumed to
remain in the gas phase \cite[e.g.][]{bergin02}. The chemical model is
based on the model of Rodgers \& Charnley \cite[]{rodgers01}, and has
been updated to incorporate recent laboratory measurements of the
dissociative recombination channels for several ions, including
CH$_3$OH$_2^+$
\cite[]{geppert06,geppert04a,geppert04b,geppert05,kalhori02}.

The initial molecular abundances in the newly-formed clump are
determined by those in the interclump medium, which are themselves
controlled by both (i) the chemistry in these regions, and (ii)
whether the material previously passed through a dense phase. This
latter effect may be important, since if most clumps eventually
dissipate, recycling dense cloud material back into the surrounding
gas, the abundances in the interclump medium retain a memory of the
higher density phase \cite[cf.][]{charnley88}. We assume that the
material in the interclump medium is predominantly molecular, and that
$\approx 99$\% of the carbon and nitrogen incorporated into a clump is
initially present as CO and N$_2$, with $\approx 1$\% in neutral
atomic form. The oxygen not bound up in CO is assumed to be atomic, as
is all the initial sulphur. To account for depletion, the gas-phase
sulphur abundance is assumed to be $10^{-8}$ \cite[]{ruffle99}, and we
assume complete depletion of all metals. The initial fractional ionization is $1.5\times10^{-7}$.

\subsection{Results}

Based on the relative binding energies of species thought to
be present on interstellar grains \cite[]{aikawa97}, we consider three types of clumps:
\begin{itemize}
\item{Type I.} No sublimation of any ice species. The initial gas-phase
  composition is equal to that of the interclump medium.
\item{Type II.} Sublimation of the most volatile ice species. We assume
  injection of CO$_2$ and H$_2$S with abundances of $10^{-5}$ and
  $10^{-7}$ respectively.
\item{Type III.} Sublimation of tightly-bound species. We inject
  CO$_2$, CH$_3$OH, and H$_2$O with abundances of $10^{-5}$,
  $2\times10^{-8}$, and $3\times10^{-5}$ respectively.
\end{itemize}
The abundances of sublimated CO$_2$ and H$_2$O are based on
observations of interstellar ices in dark clouds \cite[]{knez05}, and
the \meth\ abundance is chosen to match our derived column densities
(see section 2). H$_2$S has not been detected in interstellar ices,
but based on the fact that many of the most abundant ice species are
simple hydrides, it is likely to be present; our assumed abundance 
is consistent with the observed upper limits \cite{charnley97}. Hydrogen sulphide has a
low sublimation enthalpy \cite[]{aikawa97}, and so it will be released
along with CO$_2$ in type II cores. We neglect H$_2$S injection in
type III cores in order to distinguish between the amount of SO that
can be formed from H$_2$S, and the amount that can be formed by simply
boosting the OH abundance (as will occur when water is injected).

\begin{figure*}[t]
\includegraphics[height=15cm]{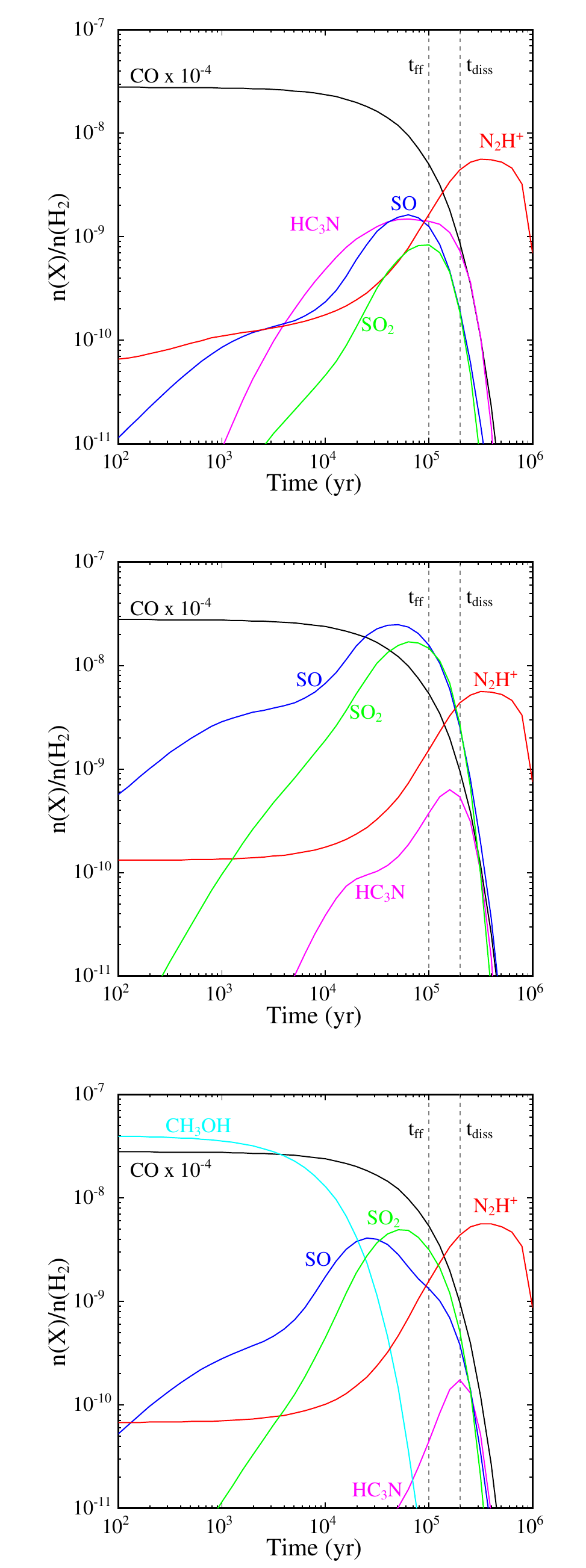}
\label{fig:model} 
\caption{Abundances versus time for selected species. The vertical
  dotted lines mark the time-scales for gravitational collapse ($t_{\rm
    ff}$) and clump dissipation ($t_{\rm diss}$). Top: type I cores
  (no ice sublimation). Middle: type II cores (CO$_2$ and H$_2$S
  sublimation). Bottom: type III cores (CO$_2$, H$_2$O, and
  \meth\ sublimation).}
\end{figure*}

Figure~\ref{fig:model} shows the time evolution of each of the four
observed species -- CO, SO, \hctn, and \meth\ -- in each clump. Also
shown are \nthp\ and SO$_2$\@. The free-fall time, $t_{\rm ff}$, and
the clump dissipation time, $t_{\rm diss}$ are marked on each
panel. In all cases, molecules are depleted from the gas on a
freeze-out timescale of a few $\times10^5$~yr. For type I clumps, SO
and HC$_3$N are produced by chemical reactions in the dense gas,
peaking between $5\times10^4$ and $10^5$~yr. SO is formed via the
reaction of atomic S with OH, and \hctn\ from ${\rm CN +
  C_2H_2}$\@. Type I clumps younger than $\sim 10^5$~yr can account
for regions with both SO and \hctn\ emission but no methanol. At
longer times, SO is destroyed faster than \hctn, so long-lived clumps
of this type are able to explain those regions which are \hctn-rich
but SO-poor. In type II clumps, the injection of H$_2$S significantly
enhances the amount of SO produced in the gas. Compared to type I
clumps, the \hctn\ peak is slightly reduced, and shifted to later
times. This is due to the reduced hydrocarbon abundances which result
from the fact that the injected CO$_2$ drives down the C$^+$
abundance. Type II cores can explain regions with large SO abundances,
together with small \hctn\ abundances and no \meth. Type III clumps
are obviously the only ones which can explain \meth-rich clumps, but
we find that the methanol only survives for $\sim 10^4$ yr before it
is destroyed by protonation followed by dissociative
recombination. The \hctn\ in these clumps only peaks {\it after} all
the \meth\ has been lost, which naturally explains the
anti-correlation between methanol and cyanoacetylene. The SO
abundances in type III clumps are intermediate between those of types
I and II, and peaks occur at around $2\times 10^4$~yr. At this time, much of
the initial \meth\ has been destroyed, and no significant amounts of
\hctn\ have been formed. Therefore, type III cores younger than $\sim
10^4$~yr will contain both \meth\ and SO, whereas slightly older cores
will only contain SO\@. Because the injected water leads to an
increased OH abundance in the gas, type III cores eventually evolve to
a state with ${\rm SO_2/SO > 1}$, so the SO/SO$_2$ ratio may
potentially be used to place more firm constraints on the ages of
these clumps \cite[c.f.][]{charnley97,buckle03}.


\subsection{Application to Observations}

We can compare our model results with the observational
dataset, to try to place constraints on the ages and the dynamical
states of specific objects. Considering the general trends described
in section 3.2, we find that our model naturally accounts for the
\meth--\hctn\ anti-correlation. Clumps with \meth\ emission must have
ages of $t \ltsim 10^4$ yr, and those with no SO emission must be even
younger ($t\ltsim 5\times10^3$~yr). Because these ages are much less
than those associated with the dynamical evolution of the clumps, the
dynamical state of methanol clumps cannot be constrained from the
observations. As these type III clumps evolve, they eventually lose
all the initial methanol; for $3\times10^4 \ltsim t \ltsim 10^5$~yr, they will
appear only in SO, after which time the \hctn\ abundance rises. The
evolution of type III clumps provides a simple explanation for the
observed transition from pure-\meth\ to mixed-\meth-SO to pure-SO gas
in different velocity channels. Clumps with both SO and
\hctn\ emission could be old type II or III ($t > 10^5$~yr), or
possibly younger type I clumps. Regions containing only
\hctn\ emission are likely to be old type I clumps, with ages $t \sim
2\times10^5~{\rm yr} \sim t_{\rm diss}$. As these latter clumps have ages
comparable to t$_{\rm diss}$, it is probable that pure HC$_3$N
clumps represent long-lived, stable clumps: gravitationally bound
against dissipation, but not collapsing. If so, then some of these regions should
also be associated with enhanced \nthp\ abundances, as we in fact see
in L1512.

We now briefly consider each source in turn:

\subsubsection*{L1512}

The strong \meth\ and SO clump in the northwest must have an age of $\sim
10^4$~yr. In contrast, the SO cores toward the southeast have only weak
methanol emission and must be older. This SO emission surrounds the
\hctn\ peak which must be even older, as no SO is seen at this
position, suggesting an age of $\sim 2\times10^5$~yr. This is
consistent with the CO depletion and \nthp\ emission also seen toward
this position \cite[]{caselli02}

\subsubsection*{TMC-1C}

The methanol clumps in this source have only weak SO emission, and so
must be $t\ltsim 5000$ yr. The clumpy ring of material to the southeast has
little or no methanol, but SO and \hctn\ are somewhat co-mingled. This
is best explained by type I clumps, implying that the mechanism which
formed the clumpy ring was not energetic enough to cause significant
desorption of grain mantles.

\subsubsection*{L1262}

In this source, the channel velocity maps at 4.12 km~s$^{-1}$ show a
clumpy ring of SO emission surrounding the \hctn\ peak. As is the case
in L1512 these may represent evolved type III clumps which have lost
the original methanol. The central \hctn\ clump will be older ($t
\gtsim 10^5$~yr) than the surrounding SO clumps ($3\times10^4 < t <
10^5$~yr). Both CO clumps in this source display the kinematical sequence 
of SO followed by SO plus \meth, followed by only \meth, that is typical
of the time evolution of type II clumps.

\subsubsection*{Per 7}

Again, we see a transition in velocity space from pure SO to pure \meth\ via
mixed gas. As in L1262, this indicates the evolution of clumps where \meth\
ice is liberated along the line of sight. The \hctn\ peak toward the northeast
 is associated with weak SO, and may
represent evolved gas. However, the peak does not appear to be correlated
with the \nthp\ peak or with significant CO depletion.

\subsubsection*{L1389}

Again, we see  SO-rich gas surrounding a \hctn-rich region.
As in L1512, this suggests an older, stable clump in the center surrounded
by regions in which clump formation and/or ice mantle sublimation
occurred more recently. This source also shows a clear \hctn-\meth\
anti-correlation.

\subsubsection*{L1251E}

SO is only visible in this map toward the north and east, and appears
to be associated with some \meth, implying an age for this gas of
$\sim 10^4$~yr. Towards the south and west, \meth\ and \hctn\ clumps
are prevalent yet SO is absent, which requires the presence of both young
and old cores. This is the only area in which we see co-mingling 
of \hctn\ and \meth\ emission -- although the peak positions are offset --
and is the only region whose chemistry is not readily explained by
our simple model.


\section{Conclusions}

We have mapped several prestellar and protostellar cores in \meth,
\hctn, SO and \CeiO. We find that the emission from these molecules
traces clumpy structure across all of the cores. The molecular clumps
show striking differences in the location and velocity of emission
peaks between molecules.

We see a similar degree of chemical differentiation
in all of the cores that we have observed, despite the differences in
known star forming activity.   The  \hctn\ clumps are generally anti-correlated with N$_2$H$^+$ peaks, and with \meth\ clumps.  In  L1512,  a  \hctn\ clump does appear along with a  N$_2$H$^+$ peak, perhaps indicative of ongoing molecular depletion. 
 \meth\ clumps and SO clumps
also show distinct emission peaks. \CeiO\ clump peaks are not
generally correlated with emission peaks in any of the other molecules
that we observed. These morphological differences in molecular peaks
towards sources of different evolutionary stage suggest that depletion
is important, but that other processes must also be driving the
chemical differentiation that we see.

Our observations suggest there is a kinematic sequence through the
clumps of different molecular species within the core. There is a well
defined emission sequence in velocity space where we see only clumps
of SO, then clumps where SO and \meth\ are intermingled, and then a
region with \meth\ only. This may indicate that we are looking down a
cylinder, or filament.

We have investigated the origin of the chemical differentiation using
a simple chemical model, in which clumps form rapidly and in which
sublimation of ice mantles can occur as the clump is formed.  Despite
its simplicity, this model is able to account for a wide variety of
the chemical differentiation observed in our sources.
For clumps in which \meth\ is injected, the methanol survives for
$\sim 10^4$~yr, so the observed \meth-rich clumps are young compared
to the time-scale for dynamical evolution. In these clumps, \hctn\ is
not produced until all the \meth\ has been destroyed, which explains
the observed anti-correlation of methanol and cyanoacetylene. The
model also predicts some some regions will have ${\rm \hctn/SO} > 1$
whereas others will have ${\rm \hctn/SO} < 1$, depending on the age of
the source and/or the degree of mantle sublimation. \hctn\ peaks at
late times, and the observed regions with only \hctn\ emission are
most likely tracing older clumps which are dynamically stable against
both collapse and dissipation.  Such  regions should also be associated
with marked CO depletion and elevated \nthp\ abundances.

In conclusion, it appears that molecular clouds containing protostars
at different epochs of star formation exhibit similar trends in their
chemical spatial differentiation. The origin of this differentiation
can be understood through simple models of the molecular gas-grain
interaction, and suggests that periodic removal of ice mantles is
occurring in dark clouds.  Mantle disruption and dark cloud chemistry
are probably closely connected to the physics of clump evolution,
dynamical-chemical models incorporating of these processes are
currently being developed \cite{rodgers05}.


\section*{Acknowledgements}
 SBC and SDR acknowledge  support  by the NASA Goddard Center
 for Astrobiology and the NASA Long Term Space Astrophysics Program,
 with partial support from the NASA Origins of Solar Systems Program.

\begin{sidewaystable} 
\scriptsize
\centering
\caption{\label{t_fits}Summary of  Onsala  Observations and Data Analysis}
\begin{tabular}{@{}c@{}c@{ }c@{ }c@{ }c@{ }c@{}c@{ }c@{ }c@{ }c@{ }c@{$~~~$}c@{ }c@{ }c@{ }c@{ }c@{}c@{ }c@{ }c@{ }c@{ }c@{ }l@{ }l}
\noalign{\smallskip}
\hline
\hline
\noalign{\smallskip}
Source &Offset &   \multicolumn{4}{c}{C$^{18}$O}   &{~~~}&  \multicolumn{4}{c}{CH$_3$OH }  &{~~~}&  
\multicolumn{4}{c}{HC$_3$N }  &{~~~}&  \multicolumn{4}{c}{SO } &  \\
\noalign{\smallskip}
\cline{3-6}
\cline{8-11}
\cline{13-16}
\cline{18-21}
\noalign{\smallskip}
& ($\arcsec$)  &  $\int  T_{\rm A}^* dv$   & $\Delta v$ & $V_{\rm LSR}$ &$N^\dagger$ &&
  $\int  T_{\rm A}^* dv$   & $\Delta v$ & $V_{\rm LSR}$ &$N^\dagger$ &&
  $\int  T_{\rm A}^* dv$   & $\Delta v$ & $V_{\rm LSR}$ &$N^\dagger$ &&
  $\int  T_{\rm A}^* dv$   & $\Delta v$ & $V_{\rm LSR}$ &$N^\dagger$ &&\\
 &       &  (K\kms)            & (\kms) & (\kms)   & ($ \rm 10^{15} cm^{-2}$)  &   &  
              (K\kms)            & (\kms) & (\kms)   & ($ \rm 10^{14} cm^{-2}$)  &   &  
             (K\kms)            & (\kms) & (\kms)   & ($ \rm 10^{12} cm^{-2}$)  &   &  
             (K\kms)            & (\kms) & (\kms)   & ($\rm  10^{13} cm^{-2}$)      \\ 
\noalign{\smallskip}    
\hline
Per7 &&&&&&&&&&&&&&&&&&\\
&-90,-60&1.77(.07)&0.85(.04)&6.26(.02)&5.43&&0.31(.03)&0.52(.03)&6.44(.02)&2.82&&0.13(.03)&0.64(.17)&6.38(.07)&0.96&&0.93(.07)&0.95(.09)&6.30(.03)&2.32&\\
&-60,0&1.80(.07)&0.72(.04)&6.32(.01)&5.51&&0.37(.03)&0.57(.03)&6.53(.02)&3.41&&$<$0.01&..&..&$<$0.09&&0.75(.07)&0.91(.11)&6.41(.04)&1.86&\\
&-30,30&2.32(.07)&0.87(.03)&6.52(.01)&7.11&&0.22(.02)&0.49(.03)&6.68(.02)&2.04&&0.09(.01)&0.20(.03)&6.81(.01)&0.68&&0.83(.06)&0.68(.06)&6.53(.03)&2.07&\\
&30,0&1.96(.06)&0.75(.03)&6.53(.01)&6.01&&0.25(.02)&0.50(.03)&6.68(.02)&2.26&&$<$0.01&..&..&$<$0.09&&0.53(.07)&0.61(.10)&6.61(.03)&1.31&\\
&30,30&2.30(.06)&0.81(.03)&6.54(.01)&7.03&&0.17(.03)&0.44(.03)&6.66(.03)&1.60&&0.11(.01)&0.26(.03&6.72(.02)&0.79&&0.84(.07)&0.81(.08)&6.55(.03)&2.10&\\
&90,90&2.17(.06)&0.78(.02)&6.62(.01)&6.64&&0.43(.04)&0.71(.04)&6.69(.00)&3.93&&$<$0.02&..&..&$<$0.11&&0.68(.06)&0.64(.07)&6.47(.03)&1.70&\\
L1389&&&&&&&&&&&&&&&&&\\
&-60,-30 &0.17(.03)&0.37(.06)&-4.81(.04)&0.51&&0.11(.03)&0.37(.09)&-4.60(.05)&1.00&&0.15(.04)&0.61(.22)&-4.81(.08)&1.11&&$<$0.02&..&..&$<$0.04&\\
&-30,-30 &0.33(.04)&0.40(.05)&-4.78(.03)&1.01&&0.14(.02)&0.29(.02)&-4.64(.01)&1.25&&0.27(.02)&0.34(.03)&-4.72(.01)&1.91&&0.19(.03)&0.29(.05)&-4.85(.02)&0.47&\\
&-30,0   &0.45(.05)&0.59(.07)&-4.81(.04)&1.38&&0.11(.03)&0.48(.08)&-4.69(.05)&1.05&&0.29(.02)&0.33(.02)&-4.67(.01)&2.07&&0.25(.03)&0.36(.04)&-4.85(.02)&0.62&\\
TMC-1C&&&&&&&&&&&&&&&&&\\
&-90,120 &1.50(.09)&0.43(.03)&5.07(.01)&4.59&&0.29(.03)&0.25(.02)&5.28(.01)&2.64&&$<$0.03&..&..&$<$0.23&&0.50(.08)&0.85(.14)&5.31(.0)8&1.24&\\
&-60,60  &1.38(.09)&0.42(.03)&5.12(.01)&4.21&&0.50(.03)&0.33(.01)&5.33(.01)&4.58&&0.15(.02)&0.22(.03)&5.25(.02)&1.06&&0.32(.04)&0.29(.04)&5.13(.02)&0.79&\\
&-30,-60 &1.48(.11)&0.47(.05)&5.24(.02)&4.54&&0.23(.04)&0.30(.03)&5.38(.02)&2.16&&0.33(.03)&0.26(.03)&5.31(.01)&2.35&&0.25(.04)&0.17(.03)&5.15(.01)&0.63&\\
&-30,30  &1.30(.08)&0.33(.02)&5.16(.01)&3.97&&0.48(.03)&0.32(.01)&5.33(.01)&4.42&&0.31(.03)&0.21(.02)&5.24(.01)&2.22&&0.45(.06)&0.44(.08)&5.22(.03)&1.13&\\
&30,-30  &1.35(.08)&0.37(.02)&5.16(.01)&4.12&&0.30(.01)&0.30(.08)&5.30(.08)&2.79&&0.24(.03)&0.26(.05)&5.08(.02)&1.72&&0.38(.05)&0.32(.05)&5.12(.02)&0.94&\\
L1512&&&&&&&&&&&&&&&&&&\\
&-60,-30 &0.65(.04)&0.31(.02)&7.02(.01)&2.00&&0.14(.02)&0.29(.03)&7.21(.02)&1.25&&0.19(.03)&0.21(.04)&7.07(.02)&1.33&&0.14(.03)&0.28(.08)&7.09(.04)&0.34&\\
&-60,120 &0.38(.04)&0.25(.03)&6.86(.01)&1.18&&0.35(.02)&0.31(.01)&6.97(.01)&3.20&&0.09(.02)&0.11(.02)&7.59(.01)&0.67&&0.42(.03)&0.31(.03)&6.85(.0)1&1.05&\\
&-30,-30 &0.42(.04)&0.22(.0)2&7.02(.01)&1.29&&0.16(.02)&0.25(.01)&7.21(.01)&1.44&&0.29(.02)&0.17(.02)&7.08(.01)&2.05&&0.57(.07)&1.13(.17)&7.05(.0)7&1.43&\\
&-30,0   &0.42(.03)&0.22(.02)&7.08(.01)&1.28&&0.18(.02)&0.25(.02)&7.21(.01)&1.67&&0.48(.03)&0.24(.02)&7.09(.01)&3.41&&$<$0.03&..&..&$<$0.07&\\
L1251E&&&&&&&&&&&&&&&&&\\
&-210,0  &2.63(.17)&1.41(.11)&-3.81(.05)&8.06&&0.95(.05)&1.24(.03)&-3.88(.03)&8.70&&0.73(.05)&0.97(.08)&-3.67(.03)&5.27&&$<$0.02&..&..&$<$0.06&\\
&-180,0  &1.71(.16)&1.40(.15)&-3.76(.07)&5.23&&0.95(.07)&1.84(.08)&-3.97(.05)&8.72&&0.58(.05)&0.76(.09)&-3.40(.03)&4.17&&$<$0.03&..&..&$<$0.07&\\
&-150,0  &1.58(.15)&1.25(.12)&-3.72(.06)&4.83&&0.77(.05)&1.11(.05)&-3.68(.04)&7.04&&0.85(.05)&0.59(.04)&-3.43(.01)&6.07&&0.80(.09)&1.46(.19)&-3.77(.0)8&1.99&\\
&-120,120&0.96(.08)&0.82(.08)&-4.38(.03)&2.93&&0.57(.07)&1.18(.09)&-4.16(.07)&5.22&&$<$0.02&..&..&$<$0.16&&2.04(.08&1.08(.05)&-4.36(.02)&5.07&\\
&-90,60  &1.72(.10)&1.02(.07)&-4.16(.03)&5.28&&1.10(.08)&1.73(.08)&-4.21(.06)&10.05&&0.17(.03)&0.27(.05)&-3.77(.02)&1.24&&1.39(.08)&1.22(.09)&-4.67(.04)&3.47&\\
L1262&&&&&&&&&&&&&&&&&&\\
&0,0     &1.01(.07)&0.51(.04)&3.82(.02)&3.08&&0.43(.02)&0.39(.01)&3.98(.01)&3.93&&0.19(.01)&0.40(.04)&3.89(.01)&1.36&&0.56(.04)&0.51(.04)&3.77(.02)&1.40&\\
&0,30    &0.86(.07)&0.58(.06)&3.85(.02)&2.64&&0.62(.03)&0.51(.01)&4.02(.01)&5.72&&0.28(.02)&0.38(.03)&3.85(.02)&1.99&&1.10(.05)&0.64(.03)&3.85(.01)&2.74&\\
&30,-30  &0.64(.06)&0.45(.06)&3.96(.02)&1.95&&0.28(.02)&0.47(.02)&4.11(.02)&2.57&&0.20(.02)&0.35(.06)&4.04(.02)&1.44&&0.42(.05)&0.48(.07)&4.04(.03)&1.05&\\
&60,-30  &1.11(.08)&0.74(.07)&4.10(.03)&3.40&&0.42(.03)&0.60(.03)&4.33(.02)&3.90&&0.30(.03)&0.53(.05)&4.06(.02)&2.16&&0.68(.05)&0.62(.05)&4.08(.02)&1.70&\\
\noalign{\smallskip} 
\hline 
\noalign{\smallskip} 
\end{tabular} 
\begin{flushleft} 
 \footnotesize 
\noindent  
NOTE: The error estimate in parenthesis is the 1$\sigma$ error estimate.
Upper limits are 3$\sigma$. For column densities, upper limits
are based on $T_{\rm A}^*$$\Delta${\it v}, where the value of $\Delta${\it v} is the channel
width, and $T_{\rm A}^*$ is taken to be 3 $\sigma$.\\ 
{$^\dagger$ see text for details} 
\end{flushleft} 
\end{sidewaystable}

\end{document}